\documentstyle[preprint,aps]{revtex}

\newcommand{\unit}{\hbox{$\hat{\bf n}$}}
\newcommand{\pol}{\hbox{$\hat{\bf e}$}}
\newcommand{\rv}{{\bf r}}
\newcommand{\Ev}{{\bf E}}
\newcommand{\Ec}{{\cal E}}
\newcommand{\dv}{{\bf d}}

\newcommand{\Dcv}{\hbox{\boldmath$\cal D$}}

\newcommand{\kappav}{\hbox{\boldmath$\kappa$}}

\newcommand{\skv}{{\hbox{\bf\scriptsize k}}}

\newcommand{\Dkappav}{\Delta\hbox{\boldmath$\kappa$}}
\newcommand{\sDkappav}{\Delta\hbox{\boldmath$\scriptstyle\kappa$}}

\newcommand{\kv}{{\bf k}}

\newcommand{\beq}{\begin{equation}}
\newcommand{\eeq}{\end{equation}}
\newcommand{\bea}{\begin{eqnarray}}
\newcommand{\eea}{\end{eqnarray}}

\begin{document}
\draft
\preprint{}
\title{Phase dependent spectrum of scattered light \\from two Bose condensates}
\author{Janne Ruostekoski and Dan F. Walls}
\address{
Department of Physics, University of Auckland, Private Bag 92019, \\Auckland,
New Zealand}
\date{\today}
\maketitle
\begin{abstract}
We calculate the spectrum of the scattered light from quantum degenerate
atomic gases obeying Bose-Einstein statistics. The atoms are assumed to
occupy two ground states which are optically coupled through a common
excited state by two low intensity off-resonant light beams. In the 
presence of a Bose condensate in both ground states, the atoms may
exhibit light induced oscillations between the two condensates analogous
to the Josephson effect. The spectrum of the scattered light is calculated
in the limit of a low oscillation frequency. In the spectrum we are able
to observe qualitative features depending on the phase difference between 
the macroscopic wave functions of the two condensates. Thus, our optical scheme
could possibly be used as an experimental realization of the 
spontaneous breakdown of the U(1) gauge symmetry in the Bose-Einstein
condensation.
\end{abstract}

The Bose-Einstein condensation (BEC) has finally been observed in a weakly
interacting system with well-understood interactions by cooling and
trapping alkali metal vapors \cite{AND95,BRA95,DAV95}. In these
first experiments on BEC the condensates were probed destructively
with a short flash of laser light. Recently, Andrews {\it et al}.\
\cite{AND96} have reported a non-destructive optical
detection of a Bose condensate. This detection technique could possibly
also be used to measure the spectrum of the scattered light from the condensate,
when the driving light is
detuned far from the resonance of the optical transitions
\cite{JAV95b,GRA96,CSO96}.

In the BEC phase transition the Bose gas should acquire nontrivial
phase properties. In the traditional reasoning the condensate is given
a macroscopic wave function which acts as a complex order parameter with an
arbitrary but fixed phase \cite{FOR75}. The selection of the phase in each
experiment implicates the spontaneous breaking of the U(1) gauge symmetry.
However, even though the condensate is taken to be in a number state with 
no phase whatsoever, the condensate behaves as 
if it had a phase \cite{JAV91,JAV96,NAR96,CIR96,JAC96,WON96,CAS96,BAR96}. 
The phases of the condensates are
expected to exhibit collapses and revivals due to collisions 
\cite{WON96,WRI96}.

Most of the theoretical studies on optical response of degenerate atomic 
gases \cite{JAV95b,GRA96,CSO96,SVI90,POL91,LEW93,JAV94,MOR95,LEW96,RUO97}
are insensitive to the phase properties of the condensates. Javanainen
\cite{JAV96b} and Imamo\=glu and Kennedy \cite{IMA96} have proposed optical
schemes to detect spontaneous breakdown of the gauge symmetry in BEC.
In Ref.~{\cite{JAV96b}} two condensates in different Zeeman states are confined
in the same trap, and two phase coherent laser beams are used to drive
Raman transitions between the condensates. Amplification of one of the
light beams, because of the stimulated Raman scattering, would indicate 
phase dependent properties. In Ref.~{\cite{IMA96}}
nonlocal light scattering between two independent, spatially separated
condensates were studied. 

In this paper we consider the spectrum of the scattered light from a
quantum degenerate 
Bose-Einstein gas occupying two ground states with a common excited state.
Following Ref.~{\cite{JAV96b}} all the atoms are confined in the same trap. The
possible Bose condensates are in two different Zeeman sublevels 
$|b\rangle=|g,m\rangle$ and $|c\rangle=|g,m-2\rangle$. The state $c$
is optically coupled to the electronically excited state $|a\rangle=|e,m-1\rangle$
by the driving field $\Ev_{d2}$ having a polarization $\sigma_+$ and a dominant
frequency $\Omega_2$. Similarly,
the state $b$ is coupled to $a$ by the driving field $\Ev_{d1}$
with a polarization $\sigma_-$ and a dominant frequency $\Omega_1$.
We assume that the light fields $\Ev_{d1}$ and $\Ev_{d2}$ propagate in
the positive $z$-direction and are detuned far 
from the resonances of the corresponding atomic transitions. The light fields
are also assumed to be in the coherent states. 
In accordance with our previous work \cite{JAV95b,RUO97} we can immediately 
write down the Hamiltonian density
\begin{mathletters}
\bea
{\cal H}(\rv)&=& \psi^\dagger_{b}(\rv)
\hbox{$H_{\hbox{\scriptsize c.m.}}$}
\psi_{b}(\rv)+ \psi^\dagger_{c}(\rv)
(\hbox{$H_{\hbox{\scriptsize c.m.}}$}+\hbar\omega_{cb})\psi_{c}(\rv)
+\psi^\dagger_{a}(\rv)
(\hbox{$H_{\hbox{\scriptsize c.m.}}$}+
\hbar\omega_{ab})\psi_{a}(\rv)+ {\cal{H}}_F(\rv)\nonumber\\
&-&\left(\dv_{ba}\cdot\Ev_1(\rv)\,\psi^\dagger_{b}(\rv)\psi_{a}(\rv)+
{\rm h.c.}\right)-\left(\dv_{ca}\cdot\Ev_2(\rv)\,
\psi^\dagger_{c}(\rv)\psi_{a}(\rv)+{\rm h.c.}\right)\,,
\label{eq:HDN}
\eea
\begin{equation}
H_F = \int d^3r\,{\cal H}_F({\bf r}) =
\hbar \sum_q\omega_q^{\rule[0.3ex]{0pt}{0pt}}
a^\dagger_q a_q^{\rule[0.3ex]{0pt}{0pt}}\,.
\label{eq:HF}
\end{equation}
\end{mathletters}
The first three terms reflect the energies, internal and
center-of-mass (c.m.), of the atoms in the absence of
electromagnetic fields. The frequencies for the optical transitions 
$a\rightarrow b$ and $a\rightarrow c$ are $\omega_{ab}$ and $\omega_{ac}$
($\omega_{cb}=\omega_{ab}-\omega_{ac}$),
respectively. The last two terms are for the atom-light dipole interaction.
The dipole matrix element for the atomic transition $a\rightarrow b$ is given 
by $\dv_{ba}$. The Hamiltonian density for the free electromagnetic 
field is ${\cal H}_F$. If the light field is fully treated as a quantum 
mechanical field with its own dynamics, the Hamiltonian may also contain the 
polarization self-energy term and the dynamical light field should be the electric 
displacement instead of the electric field \cite{RUO97,COH89}. However, if the 
detunings ($\Delta_1=\Omega_1-\omega_{ab}\simeq\Delta_2=\Omega_2-\omega_{ac}$) are 
sufficiently large \cite{JAV96b}, multiple scattering is negligible and,
at first order in $1/\Delta_1$, these effects may be ignored.

We define slowly varying field operators in the Heisenberg picture by
$\tilde{\psi}_{a} = e^{i\Omega_1 t}\psi_{a}$,
$\tilde{\bf E}^+_1 = e^{i\Omega_1 t}{\bf E}^+_1$,
$\tilde{\bf E}^+_2 = e^{i\Omega_2 t}{\bf E}^+_2$, and
$\tilde{\psi}_{c} = e^{i(\Omega_1-\Omega_2) t}\psi_{c}$.
In the limit of large detuning,
the excited state field operator $\psi_a$ may be eliminated adiabatically:
\beq
\tilde{\psi}_{a}(\rv) ={1\over \hbar\Delta_1}\left(\dv_{ab}\cdot
\tilde{\bf E}^+_1(\rv)\,
\psi_{b}(\rv)+\dv_{ac}\cdot\tilde{\bf E}^+_2(\rv)\,\tilde{\psi}_{c}(\rv)
\right)\,.
\label{eq:adi}
\eeq
We insert Eq.~{(\ref{eq:adi})} into the Hamiltonian (\ref{eq:adi}) and keep
only the terms of first order in $1/\Delta_1$. If the
detuning from the two-photon resonance is given by $\delta_{cb}=
\Omega_1-\Omega_2-\omega_{cb}$, the Hamiltonian density now reads
\bea
{\cal H}&=& \psi^\dagger_{b}
\hbox{$H_{\hbox{\scriptsize c.m.}}$}
\psi_{b}+ \tilde{\psi}^\dagger_{c}(\hbox{$H_{\hbox{\scriptsize c.m.}}$}-
\hbar\delta_{cb})\tilde{\psi}_{c}+{\cal{H}}_F-{1\over\hbar\Delta_1}\left\{
\dv_{ba}\cdot\tilde{\bf E}^-_1\,\dv_{ab}\cdot\tilde{\bf E}^+_1
\,\psi^\dagger_{b}\psi_{b}\right.\nonumber\\
&+&\left.\dv_{ac}\cdot\tilde{\bf E}^+_2\,
\dv_{ca}\cdot\tilde{\bf E}^-_2\,\tilde{\psi}^\dagger_{c}
\tilde{\psi}_{c}+\left(\dv_{ba}\cdot\tilde{\bf E}^-_1\,
\dv_{ac}\cdot\tilde{\bf E}^+_2\psi^\dagger_{b}
\tilde{\psi}_{c}+{\rm h.c.}\right)\right\}\,.
\label{eq:HDN2}
\eea
Then, the electric fields may be solved and, according to Ref.~{\cite{JAV95b}},
the scattered fields are given by
\begin{mathletters}
\beq
\tilde{\bf E}^+_{s1}({\bf r}) = {1\over\hbar\Delta_1}\int d^3r'\,
{\bf K}({\bf d}_{ba};{\bf r}-{\bf r'})\,\left\{\dv_{ab}\cdot
\tilde{\bf E}^+_{d1}(\rv)\,\psi^\dagger_{b}({\bf r'})\psi_{b}({\bf r'})
+\dv_{ac}\cdot\tilde{\bf E}^+_{d2}(\rv)
\,\psi^\dagger_{b}({\bf r'})\tilde{\psi}_{c}({\bf r'})\right\}\,,
\label{eq:FEQa}
\eeq
\beq
\tilde{\bf E}^+_{s2}({\bf r}) = {1\over\hbar\Delta_1}\int d^3r'\,
{\bf K}({\bf d}_{ca};{\bf r}-{\bf r'})\,\left\{\dv_{ac}\cdot
\tilde{\bf E}^+_{d2}(\rv)\,\tilde{\psi}^\dagger_{c}({\bf r'})\tilde{\psi}_{c}
({\bf r'})+\dv_{ab}\cdot\tilde{\bf E}^+_{d1}(\rv)
\,\tilde{\psi}^\dagger_{c}({\bf r'})\psi_{b}({\bf r'})\right\}\,.
\label{eq:FEQb}
\eeq
\label{eq:FEQ}
\end{mathletters}
Here we have used the first Born approximation based on the assumption that
the incoming fields dominate inside the sample as the multiple scattering
is negligible. The kernel ${\bf K}(\Dcv;{\bf r}-{\bf r}')$ is the familiar
expression \cite{JAC75} of the positive-frequency component of the electric 
field from a monochromatic dipole with the complex amplitude $\Dcv$, given 
that the dipole resides at $\rv'$ and the field is observed at $\rv$.

We assume the driving electric fields to be plane waves
\beq
\tilde{\bf E}^+_{d1}({\bf r}) =\hbox{$1\over2$}
{\cal E}_1 \hbox{$\hat{\bf e}_-$} 
e^{i\hbox{\boldmath$\scriptstyle\kappa$}_{1}
\cdot\hbox{\bf\scriptsize r}},\quad
\tilde{\bf E}^+_{d2}({\bf r}) =\hbox{$1\over2$}
{\cal E}_2 \hbox{$\hat{\bf e}_+$} 
e^{i\hbox{\boldmath$\scriptstyle\kappa$}_{2}
\cdot\hbox{\bf\scriptsize r}}\,.
\label{eq:INF}
\eeq
We insert these into the Hamiltonian (\ref{eq:HDN2}) to obtain the Hamiltonian
for the matter field dynamics
\beq
{\cal H}_M= \psi^\dagger_{b}
(\hbox{$H_{\hbox{\scriptsize c.m.}}$}-\hbar\delta_1)
\psi_{b}+ \tilde{\psi}^\dagger_{c}(\hbox{$H_{\hbox{\scriptsize c.m.}}$}-
\hbar\delta_{cb}-\hbar\delta_2)\tilde{\psi}_{c}+\left(\hbar\kappa\psi^\dagger_{b}
\tilde{\psi}_{c}e^{-i\hbox{\boldmath$\scriptstyle\kappa$}_{12}
\cdot\hbox{\bf\scriptsize r}}+{\rm h.c.}\right)\,,
\label{eq:HDN3}
\eeq
where $\kappav_{12}=\kappav_1-\kappav_2$ is the wavevector difference of
the incoming light fields. Following the notation in Ref.~{\cite{JAV96b}}
\beq
\delta_1={|{\cal E}_1|^2d_{ab}^2\over 4\hbar^2\Delta_1},\quad 
\delta_2={|{\cal E}_2|^2d_{ac}^2\over 4\hbar^2\Delta_1},\quad \kappa=
{{\cal E}^*_1{\cal E}_2d_{ab}d_{ac}\over 4\hbar^2\Delta_1}\,,
\label{eq:para}
\eeq
The dipole matrix element $d_{ab}$ contains the reduced dipole matrix element
and the corresponding nonvanishing Clebsch-Gordan coefficient. To simplify
the algebra, we assume $\kappa$ to be real.

We assume a translationally invariant and non-interacting Bose gas as in
Ref.~{\cite{JAV96b}}. The matter field operators are given by the familiar 
plane wave representations $\psi_b (\rv t)=V^{-1/2}\sum_{\skv}
e^{i\skv\cdot\hbox{\bf\scriptsize r}}\,b_{\skv}(t)$ and
$\tilde{\psi}_c (\rv t)=V^{-1/2}\sum_{\skv}e^{i\skv
\cdot\hbox{\bf\scriptsize r}}\,\tilde{c}_{\skv}(t)$.
In the absence of light, the c.m. motion in both ground states
satisfies the dispersion relation for the frequency 
$\epsilon_{\skv}=\hbar|\kv|^2/2m$.

By assuming that $\kappav_{12}$ is very small we may neglect its contribution
to the time evolution.
The dynamics of the ground state annihilation operators $\tilde{c}_{\skv}$ and 
$b_{\skv}$ may then be easily solved from the Hamiltonian (\ref{eq:HDN3}).
With the definitions $2\bar{\delta}=\delta_{cb}-\delta_1+\delta_2$ and 
$\Omega_R=(\bar{\delta}^2+\kappa^2)^{1/2}$ the solutions are given by
\begin{mathletters}
\beq
\tilde{c}_{\skv}(t)=e^{i(\bar{{\hbox{$\scriptstyle\delta$}}}+
{\hbox{$\scriptstyle\delta$}}_1-{\epsilon_{{\hbox{\bf\scriptsize k}}}})t}
\left\{\tilde{c}_{\skv}(0)\left(\cos{\Omega_R t}
+{i\bar{\delta}\over \Omega_R}\sin{\Omega_R t}\right)-{i\kappa\over\Omega_R}
b_{\skv-\hbox{\boldmath$\scriptstyle
\kappa$}_{12}}(0)\sin{\Omega_R t}\right\}\,,
\eeq
\beq
b_{\skv}(t)=e^{i(\bar{{\hbox{$\scriptstyle\delta$}}}+
{\hbox{$\scriptstyle\delta$}}_1-{\epsilon_{{\hbox{\bf\scriptsize k}}}})t}
\left\{b_{\skv}(0)\left(\cos{\Omega_R t}
-{i\bar{\delta}\over \Omega_R}\sin{\Omega_R t}\right)-{i\kappa\over\Omega_R}
\tilde{c}_{\skv+\hbox{\boldmath$\scriptstyle
\kappa$}_{12}}(0)\sin{\Omega_R t}\right\}\,.
\eeq
\label{eq:osc}
\end{mathletters}
Before the light is switched on, the atoms in the states $b$ and $c$ are 
assumed to be uncorrelated. The driving light fields induce a coupling between the
two levels. In the presence of Bose condensates in the
ground states, the coupling between the two
condensates is analogous to the coherent tunneling of Cooper pairs in a Josephson
junction \cite{JAV86,JAV96b}.

According to Ref.~{\cite{JAV95b}}, the spectrum is obtained by
calculating correlation functions for the matter field operators. In particular,
in the presence of the condensate the expectation values for single operators
are nonvanishing
$\langle\psi_c(0)\rangle=(N_{c}/V)^{1/2}\,e^{i\varphi_c}$ and
$\langle\psi_b(0)\rangle=(N_{b}/V)^{1/2}\,e^{i\varphi_b}$.
We have assumed that the condensates in the states $c$ and $b$ 
have the expectation values
for the number operators $N_{c}$ and $N_{b}$, respectively. 
The condensate phases $\varphi_c$ and
$\varphi_b$ are random variables and are fixed in each experiment.

From the equations (\ref{eq:FEQ}) and (\ref{eq:osc}) we immediately see that
the spectrum of the scattered light is time-dependent. However,
with the measurement times short enough compared to the characteristic time scale
of the changes in the spectrum, the
spectrum may be considered steady. Evidently, we do not want too short measurement
times which would wipe out qualitative features from the spectrum. According to
Ref.~{\cite{JAV95b}}, the most relevant choices for the characteristic
frequency scales in the spectrum are the effective recoil frequencies $\omega_{Ri}
=\hbar(\Dkappav_i)^2/2m$, where $\Dkappav_i=\Omega_i\unit/c-\kappav_i$ is the 
change of the wave vector of the light field $\Ev_i$ upon scattering and
$\unit$ is a unit vector pointing into the direction of the scattered light.

We present a set of assumptions based on the limit of a weak
coupling between the ground states to obtain a stable spectrum during short 
measurement times.
We assume that the oscillation frequency $\Omega_R$ is small,
$\omega_{Ri}\gg\Omega_R$, or $\omega_{Ri}\gg\bar{\delta}$ and $\omega_{Ri}\gg
\kappa$. The latter indicates $\Delta_1\omega_{Ri}\gg{\cal R}_{1}
{\cal R}_{2}$, where ${\cal R}_{1}=d_{ab}\Ec_1/2\hbar$ and 
${\cal R}_{2}=d_{ac}\Ec_2/2\hbar$ are the Rabi
flopping frequencies for the electronic transitions $b\rightarrow a$ and 
$c\rightarrow a$, respectively. 
The quantities ${\omega_{Ri}}$ are expected to be in the
neighborhood of atomic recoil frequencies,
${\omega_{Ri}} \sim \epsilon_R \sim 100$~kHz as a rule of thumb \cite{JAV95b}.
The detuning $\Delta_1$ may be chosen large keeping in mind that 
the intensity of the scattered light scales as $1/\Delta_1^2$.
If the detuning from the two photon resonance becomes small, the effective
linewidth $\bar{\gamma}$ of the transition $c\rightarrow b$ may have an effect.
However, it may be shown to be proportional to $\Delta_1^{-2}$ or smaller.
Thus, we can also safely assume $\omega_{Ri}\gg\bar{\gamma}$. 
The requirement, for the measurement times $T$ to be short enough, 
sets the conditions $\kappa T\ll 1$ or $\Delta_1\gg T{\cal R}_{1}{\cal R}_{2}$
and $\bar{\delta}T\ll 1$.
With these assumptions the time evolution for the atom operators from
Eq.~{(\ref{eq:HDN3})} with the short measurement times
becomes very simple: $\tilde{c}_{\skv}(t)=
\tilde{c}_{\skv}(0)\,e^{i(
{\hbox{$\scriptstyle\delta$}}_1-{\epsilon_{{\hbox{\bf\scriptsize k}}}})t}$ 
and $b_{\skv}(t)=b_{\skv}(0)\,e^{i({\hbox{$\scriptstyle\delta$}}_1
-{\epsilon_{{\hbox{\bf\scriptsize k}}}})t}$.

The two components of the scattered light $\Ev_{s1}$ Eq.~{(\ref{eq:FEQa})}, and 
$\Ev_{s2}$ Eq.~{(\ref{eq:FEQb})}, have polarizations $\sigma_-$ and $\sigma_+$,
respectively. By selecting the polarization we may detect these fields
separately. Let us first consider the spectrum of the scattered field $\Ev_{s1}$.
To detect only $\Ev_{s1}$, we need to select the polarization
$\pol=(\sqrt{2}\pol_{-1}-\tan\theta\,e^{-i\phi}\pol_z)/\sqrt{\tan^2\theta+2}$
from the scattered light, where $\theta$ is the scattering angle in the
conventional spherical representation (the angle of the scattered light
with respect to the positive $z$-axis). The polar angle
$\phi$ is the angle with respect to the $x$-axis in the $xy$-plane. This
angle has an influence on scattering processes in which the phases of
the two circularly polarized light beams are important.
The calculation of the angular distribution
of the scattered light with the given polarizations and atomic transitions is
explained in Ref.~{\cite{JAV95b}}. If we neglect the forward scattering,
immediately after switching on the light the spectrum is given by
\begin{mathletters}
\begin{equation}
{\cal S}_1(\rv;\omega)=C_1(\rv)\,[{\cal S}_1(\omega)+{\cal S}_{1b}(\omega)],\quad
C_1(\rv) = {\Omega_1^4d_{ab}^2\over 16(\tan^2\theta+2)
\pi^2\Delta_1\epsilon_0 c^3r^2}\,,
\label{eq:SPx1b}
\end{equation}
\bea
\lefteqn{{\cal S}_1(\omega) =
\delta_1\,\delta(\omega+\omega_{R1})\,N_b
(1+\bar{n}_{b\sDkappav_1}')+\delta_1\,\delta(\omega-\omega_{R1})\,N_b
\bar{n}_{b\sDkappav_1}'+\delta_2\,\delta(\omega+\omega_{R2})\,
N_c(1+\bar{n}_{b\sDkappav_2}')}\nonumber\\
&&\mbox{}+\delta_2\,\delta(\omega-\omega_{R2})\,N_b\bar{n}_{c\sDkappav_2}'
+2\kappa\cos(\Delta\varphi)\,\delta(\omega+\omega_{R1})\,
\delta(\Dkappav_1-\Dkappav_2)\,\sqrt{N_{c}N_{b}}(1+
\bar{n}_{b\sDkappav_1}')\,,
\label{eq:SPx1a}
\eea
\bea
{\cal S}_{1b}(\omega)&& =
\sum_{\hbox{\bf\scriptsize k}}\delta_1\,\delta(\omega-\epsilon_{\skv}+
\epsilon_{\skv-\sDkappav_1})\,\bar{n}_{b{\hbox{\bf\scriptsize k
}}}'(1 + \bar{n}_{b,{\hbox{\bf\scriptsize k}}-\Delta
{\hbox{\boldmath$\scriptstyle\kappa$}}_1}')\nonumber\\
&&\mbox{}+\sum_{\hbox{\bf\scriptsize k}}\delta_2\,
\delta(\omega-\epsilon_{\skv}+\epsilon_{\skv-\sDkappav_2})
\,\bar{n}_{c{\hbox{\bf\scriptsize k}}}'(1 + \bar{n}_{b,{\hbox{\bf\scriptsize k}}
-\Delta{\hbox{\boldmath$\scriptstyle\kappa$}}_2}')\,.
\label{eq:back}
\eea
\label{eq:SPx1}
\end{mathletters}
We have explicitly separated the contribution of the Bose condensates into
${\cal S}_1(\omega)$. This term is nonvanishing only, 
if there is a condensate present at least in one of the states $b$ or $c$.
The term ${\cal S}_{1b}(\omega)$ arises from scattering processes between 
the noncondensate atoms. 
The expectation values for the number operators for
the states $c$ and $b$ in the c.m. state $\kv$ are 
$\bar{n}_{c{\hbox{\bf\scriptsize k}}}=(z^{-1}e^{\beta\hbar(\epsilon_{
{\hbox{\bf\scriptsize k}}}+\omega_{cb})}-1)^{-1}$ and
$\bar{n}_{b{\hbox{\bf\scriptsize k}}}=(z^{-1}e^{\beta\hbar\epsilon_{
{\hbox{\bf\scriptsize k}}}}-1)^{-1}$. 
The primes in these terms indicate that 
the condensate states are excluded to avoid double counting. For example, 
the notation in the last term in Eq.~{(\ref{eq:SPx1a})}
$\bar{n}_{b\sDkappav_1}'$ should be interpreted
as $\Dkappav_1\neq0$. The difference
between the condensate phases is written as $\Delta\varphi=\varphi_b-\varphi_c$.
We have used the same normalization in Eq.~{(\ref{eq:SPx1})} as in 
Ref.~{\cite{JAV95b}}.

The first term in Eq.~{(\ref{eq:back})} describes a scattering process 
in which an atom in the
ground state $b$ with the c.m. state $\kv$ scatters to the c.m. state
$\kv-\Dkappav_1$ still remaining in the state $b$. 
The delta function dictates the energy conservation, which coincides with
the theory for Doppler velocimetry of atoms \cite{hetdyn} shifted by the effective
recoil frequency $\omega_{R1}$ \cite{JAV95b}. The product of the
occupation numbers indicates that the scattering is {\it enhanced} if
the final state is already occupied. The first term in Eq.~{(\ref{eq:back})}
and the first two terms in Eq.~{(\ref{eq:SPx1a})} coincides with the results
given in Ref.~{\cite{JAV95b}}, apart from the angular 
distribution of the scattered
light and possibly the Clebsch-Gordan coefficients of the atomic transitions.
In the second term in Eq.~{(\ref{eq:back})} an atom scatters from the ground 
state $c$ to the ground state $b$, while the c.m. state undergoes the change 
$\kv\rightarrow\kv-\Dkappav_2$.

The last term in Eq.~{(\ref{eq:SPx1a})} is purely a consequence of the 
broken gauge symmetry in BEC and it depends on the phase difference 
$\Delta\varphi$
between the two condensates in the states $b$ and $c$. It varies from 
measurement to measurement, because $\Delta\varphi$ is essentially fixed in each
experiment as a random number. The delta function
for momenta indicates the exact conservation of the momentum of the two scattered
photons for an ideal gas in a spatially homogeneous system.
For a weakly interacting gas in a magnetic trap, the momenta of the 
scattered photons do not need to be exactly the same.

The last term in Eq.~{(\ref{eq:SPx1a})} is strongly peaked at $\omega=-\omega_{
R1}$. With $\Dkappav_1=\Dkappav_2$ and by setting $N_b=N_c$ and $\delta_1=
\delta_2=\kappa$ the condensate part of the spectrum 
reduces to an especially simple form:
\beq
{\cal S}_1(\omega) =
2\delta_1\,\delta(\omega+\omega_{R})\,N_b
(1+\bar{n}_{b\sDkappav}')[1+\cos{(\Delta\varphi)}]+
\delta_1\,\delta(\omega-\omega_{R})\,N_b
(\bar{n}_{b\sDkappav}'+\bar{n}_{c\sDkappav}')\,.
\label{eq:newsp}
\eeq
The effect of the phase difference between the condensates is clearly
observed. The
peak at $\omega=-\omega_R$ reaches its maximum at $\Delta\varphi=0$ and 
completely vanishes at $\Delta\varphi=\pi$.

To detect only $\Ev_{s2}$, we need to select from the
scattered light the polarization
$\pol=(\sqrt{2}\pol_{+1}+\tan\theta\,e^{i\phi}\pol_z)/\sqrt{\tan^2\theta+2}$.
The scattering spectrum ${\cal S}_2(\rv\omega)$ obtained in this case is the same
as ${\cal S}_1(\rv\omega)$ given by Eq.~{(\ref{eq:SPx1})} with the following 
obvious changes: $c\leftrightarrow b$, $\Dkappav_1\leftrightarrow\Dkappav_2$,
and $\Omega_2\rightarrow\Omega_1$.

If the detector is insensitive to polarization of the scattered light,
the spectrum has contributions from ${\cal S}_1(\rv;\omega)$, 
${\cal S}_2(\rv;\omega)$,
and from the cross terms ${\cal S}_{12}(\rv;\omega)\propto\langle
\tilde{\bf E}^-_{s2}\tilde{\bf E}^+_{s1}\rangle+\langle\tilde{\bf E}^-_{s1}
\tilde{\bf E}^+_{s2}\rangle$. These cross terms depend on the phases of the
two circular polarizations and the scattered light does not have a rotational
symmetry around the $z$-axis.
Neglecting the forward scattering, the resulting spectrum, immediately after
switching on the light, reads
\begin{mathletters}
\begin{equation}
{\cal S}_1(\rv;\omega)=C'_1(\rv)\,[{\cal S}_1(\omega)+{\cal S}_{1b}(\omega)]
+C'_2(\rv)\,[{\cal S}_2(\omega)+{\cal S}_{2b}(\omega)]+
{\cal S}_{12}(\rv\omega)\,,
\eeq
\beq
C'_1(\rv) = {\Omega_1^4d_{ab}^2(2-\sin^2\theta)\over 64
\pi^2\Delta_1\epsilon_0 c^3r^2},\quad
C'_2(\rv) = {\Omega_2^4d_{ac}^2(2-\sin^2\theta)\over 64
\pi^2\Delta_1\epsilon_0 c^3r^2}\,,
\end{equation}
\beq
{\cal S}_{12}(\rv\omega)={\Omega_1^2\Omega_2^2d_{ab}d_{ac}
\sin^2\theta
\over 32\pi^2\Delta_1\epsilon_0 c^3r^2}\cos{(2\phi+\Delta\varphi)}
\sqrt{N_bN_c}\,\left(\delta_1\,\delta(\omega
-\omega_{R1})\,\bar{n}_{b\sDkappav_1}'+\delta_2\,\delta(
\omega-\omega_{R2})\,\bar{n}_{c\sDkappav_2}'\right)
\label{eq:holespecc}
\eeq
\label{eq:holespec}
\end{mathletters}
The spectra ${\cal S}_1(\omega)$ and ${\cal S}_{1b}(\omega)$ are given by
Eq.~{(\ref{eq:SPx1})} and
the spectra ${\cal S}_2(\omega)$ and ${\cal S}_{2b}(\omega)$ may be obtained 
from the same expressions
with the previously explained changes. The new term {(\ref{eq:holespecc})}
also is purely a consequence of the phase difference between the two condensates.
By setting $\Dkappav_1=\Dkappav_2$, $N_b=N_c$, $d_{ab}=d_{ac}$, $\delta_1=
\delta_2=\kappa$, and the scattering angle $\theta=\pi/2$ the total 
scattered spectrum may be expressed in the simple form:
${\cal S}(\rv;\omega)=C'_1(\rv)\,[{\cal S}(\omega)+{\cal S}_{b}(\omega)]$, where
\begin{mathletters}
\bea
{\cal S}(\omega)&=&2\delta_1\,\delta(\omega+\omega_{R})\,N_b
(2+\bar{n}_{b\sDkappav}'+\bar{n}_{c\sDkappav}')[1+\cos{(\Delta\varphi)}]
\nonumber\\&&\mbox{}+2\delta_1\,\delta(\omega-\omega_{R})\,N_b
(\bar{n}_{b\sDkappav}'+\bar{n}_{c\sDkappav}')[1+\cos{(2\phi+\Delta\varphi)}]\,,
\label{eq:kokoa}
\eea
\beq
{\cal S}_{b}(\omega)=\sum_{\alpha,\beta} 
\sum_{\hbox{\bf\scriptsize k}}\delta_1\,
\delta(\omega-\epsilon_{\skv}+
\epsilon_{\skv-\sDkappav})\,\bar{n}_{\alpha{\hbox{\bf\scriptsize k
}}}'(1 + \bar{n}_{\beta,{\hbox{\bf\scriptsize k}}-\Delta
{\hbox{\boldmath$\scriptstyle\kappa$}}}')\,.
\label{eq:kokob}
\eeq
\label{eq:koko}
\end{mathletters}
Here the summation over $\alpha$ and $\beta$ denotes the summation over
the ground states $b$ and $c$. The phase difference between the condensates
again has clearly observable effects on the spectrum. The two condensate
peaks in Eq.~{(\ref{eq:kokoa})} oscillate 
as a function of the phase difference between the two condensates. However,
these oscillations depend on the observation point of the scattered light
(the angle $\phi$ in the $xy$-plane). The detection of the scattered light at
several different points in space (with different values of $\phi$)
may provide a scheme
to determine the phase difference between the condensates very accurately.

The integration of Eq.~{(\ref{eq:kokob})} is straightforward in the
continuum limit and the results are given in Ref.~{\cite{JAV95b}}.
For simplicity, we choose $\omega_{cb}=0$, so that the statistical
distributions of the states $b$ and $c$ are identical. We give representative
spectra of Eq.~{(\ref{eq:koko})} in Fig.~{\ref{fig:1}} with different 
values of $\Delta\varphi$. We choose the detection point at $\theta=\pi/2$
and $\phi=\pi/2$. To simplify further, we set
$\omega_R=\omega_D=[k_BT(\Dkappav)^2/m]^{1/2}$, where
$\omega_D$ is the effective Doppler width corresponding to the change
of the wave vector $\Dkappav$. For $^{87}$Rb atoms this corresponds 
approximately to the temperature 270 nK.
We remove the integrable singularities
of Eqs.~{(\ref{eq:kokoa})} and {(\ref{eq:kokob})} by convoluting the
spectra with a Gaussian whose variance is 0.01 $\omega_R$.
It is assumed that a fraction 0.8 of the atoms
in both ground states are in the Bose condensate.

At $\Delta\varphi=0$ the peak at $\omega=-\omega_R$ has its maximum height.
The contribution of the scattering processes into 
the condensates completely vanishes in the spectrum.
The peak at $\omega=\omega_R$ has its minimum height.
At $\Delta\varphi=\pi /2$ the cosine terms in Eq.~{(\ref{eq:kokoa})} vanish.
At $\Delta\varphi=\pi$
the contribution of the scattering processes out of the condensates
vanishes in the spectrum. 
The peak at $\omega=-\omega_R$ reaches its minimum, while the peak at 
$\omega=\omega_R$ is in its maximum.

The spectra {(\ref{eq:SPx1})} and {(\ref{eq:holespec})} were calculated
immediately after switching the light on. The expectation value of the
atom number in the two ground states oscillates with the frequency
$2\Omega_R$, see Eq.~{(\ref{eq:osc})}. The atom numbers of the condensates
oscillate even if the number of atoms in each condensate is initially equal,
analogous to the Josephson effect \cite{JAV86,JAV96b}. With 
$\bar{\delta}\simeq 0$ and $\kappav_{12}\simeq 0$ the expectation values
for the number operators in the two condensates would approximately
interchange in the time period of $\pi / 2\kappa$. 

We have calculated the spectrum for an ideal, noninteracting gas. In a magnetic
trap and in the case of a weakly interacting gas the momentum distribution
is broadened, due to both the uncertainty principle and the
interparticle interactions. However, the qualitative features in the spectrum
are still well recognisable \cite{GRA96,CSO96}. If we wanted to consider 
Bose condensates in arbitrary momentum states, 
we would evidently have to abandon the assumption of translational invariance 
in space and use some other eigenfunction
basis than the plane wave basis for matter field operators.

In conclusion, the spectrum of the scattered light
from two Bose condensates in a trap may show an unambigious signature
of the broken gauge symmetry in BEC. The phase difference of
the macroscopic wave functions for the condensates has clearly observable
effects on the spectrum and its value may possibly be
determined very accurately from the oscillating heights of the peaks. 
Recently, Myatt
{\it et al}. \cite{MYA96} have managed to produce two overlapping Bose condensates
in two different angular momentum states of $^{87}$Rb. The condensates were
created using nearly lossless sympathetic cooling of one state via thermal
contact with the other evaporatively cooled state. The two states
$|F=1,m=-1\rangle$ and $|F=2,m=2\rangle$ cannot be coupled by two photon
transitions,
as was assumed in our calculations, but the generalization for the three
photon case should be rather straightforward.

\subsection*{Acknowledgements}
This work was supported by the Marsden Fund of the Royal Society of
New Zealand, The University of Auckland Research Fund and The New Zealand
Lottery Grants Board.

\begin{figure}
\caption{Spectra of light scattered from a Bose gas occupying 
two ground states with various values
of the phase difference $\Delta\varphi$ between the two condensates.
The scattered light is measured at $\theta=\pi/2$ and $\phi=\pi/2$.
A fraction 0.8 of the atoms in both ground states are in the Bose condensate.
For the solid line $\Delta\varphi=0$, for the dashed line $\Delta\varphi=\pi /2$,
and for the dotted line $\Delta\varphi=\pi$. The origin corresponds to the
common laser frequency $\Omega_1=\Omega_2$.
The spectra are expressed in units of the effective recoil frequency
$\omega_R$, and the effective Doppler width is chosen as
${\omega_D}$ = $\omega_R$.
To regularize integrable divergences, all computed spectra are convolved with
a Gaussian with the variance $0.01\,\omega_R$.
}
\label{fig:1}
\end{figure}

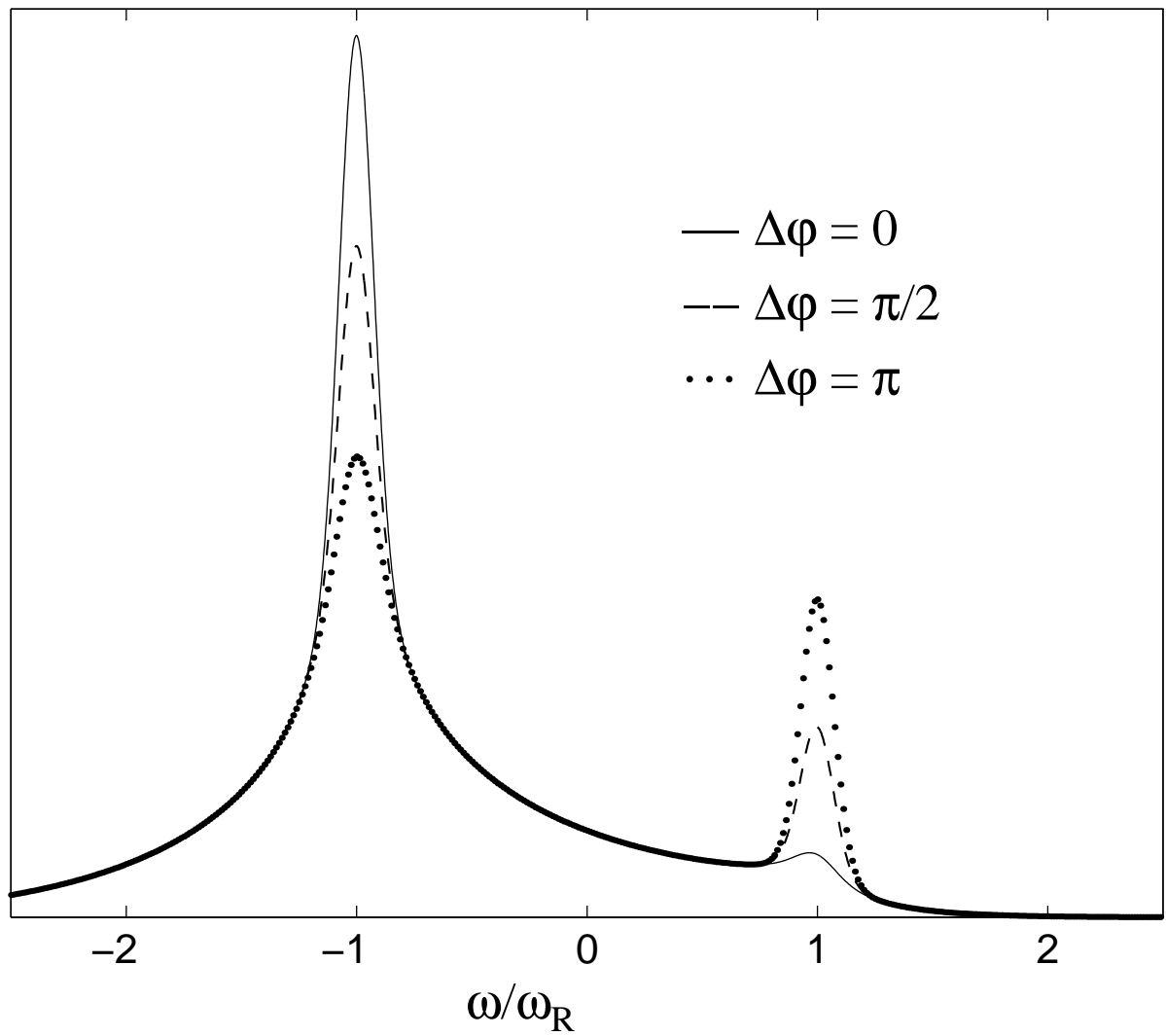
\end{document}